\documentstyle[fleqn,espcrc2,epsfig]{article}
\title{
$K\to \pi\pi$ Decay Amplitude on the Lattice
}
\author{
N.~Ishizuka
\address{
Center for Computational Physics,
University of Tsukuba, 
Tsukuba, 
Ibaraki 305-8577, 
Japan
}
}
\begin{document}
%
%---------------------------------------------------------------------------
%
\begin{abstract}
\vspace{-0.2cm}
Recent theoretical and numerical progresses of the lattice calculations
of $K\to\pi\pi$ decay amplitude are reviewed.
\end{abstract}
\maketitle
%
%---------------------------------------------------------------------------
%
\section{ Introduction }
Despite the full understanding of the fundamental theory of weak interactions,
the non-leptonic decay of hadrons still remains 
as the least understood of weak processes,
the most notable problems being the $\Delta I=1/2$ rule and
the value of $\epsilon'/\epsilon$.
The predicament originates from the difficulty
of evaluating the $K\to\pi\pi$ decay amplitudes.
In spite of recent progress of computational power,
the calculation of the decay amplitude by lattice QCD simulations
is much more difficult than those of hadron masses.
There are two difficulties in the calculations, one of which 
is the operator matching of $4$-fermion weak operators, and
the other is the problems first pointed out by Maiani and Testa~\cite{MT-nogo}.

For the $\Delta I=1/2$ decay process,
operator mixings with lower dimensional operators can occur.
Since these operators do not contribute to the physical decay amplitude,
we have to subtract the effect of these in a non-perturbative manner.
The non-perturbative subtraction is not easy numerically,
but explicit methods have been proposed
and carried out in actual lattice calculations.
The details are reviewed in Ref.~\cite{DMRSSTT}.
A recent progress is the realization that twisted mass QCD approach may 
provide a powerful tool for solving the operator mixing~\cite{tmQCD,Vladikas}. 

The most serious problem in the evaluation of the decay amplitude
was pointed out by Maiani and Testa ~\cite{MT-nogo}.
We have to calculate the decay amplitude
into a two-pion state with non-zero momenta.
However, it is non-trivial to extract it from the corresponding Green's function
$K(\vec{0}) \to \pi(\vec{p}) \pi(-\vec{p})$,
since the two-pion state with non-zero momenta is an excited state in the two-pion system.
This problem of extraction of the decay amplitude
is one of the problems pointed out by Maini and Testa,
which we shall refer to as ``MT-1'' in this article.

To avoid this problem, in most of lattice calculations,
the decay amplitudes are calculated
either at an un-physical kinematics
or through the $K\to\pi$ amplitudes and chiral perturbation theory (CHPT) 
relating these amplitudes to the physical decay amplitudes.
Using such an effective theory, however, is the cause for large uncertainties
of the lattice prediction of the decay amplitude.

Even if we succeed in overcoming the MT-1 and extract the decay amplitude
at the physical kinematics from the Green's functions, 
another problem exists.
Since lattice simulations are carried out on a finite Euclidean space-time,
the decay amplitude on the lattice does not directly give the physical one
in the infinite volume Minkowski space-time.
This problem of the relation between the lattice and the physical amplitude
is another part of the Maini--Testa problem,
which we shall call as ``MT-2'' in this article.

Recently Lellouch and L\"uscher derived a relation
between the two amplitudes~\cite{Lellouch-Luscher}.
Their derivation does not rely on effective theories.
Lin {\it et.al.} reached the same relation from a different approach, 
and extended it to general kinematics~\cite{LMST}.
They also examined the conditions for the validity of the relation.

In this article I focus on the Maini-Testa problems.
Recent theoretical and numerical progress for avoiding or solving the MT-1
will be discussed.
We also show a brief derivation of the relation by Lellouch and L\"uscher.
We refer to the reviews of weak matrix elements in recent lattice conferences
by Lellouch~\cite{Lat01_plen} and Martinelli~\cite{Lat02_plen}, 
and Ref.~\cite{DMRSSTT}
for other recent theoretical and numerical progresses.
%
%----------------------------------------------------------------------------
%
\section{ The MT-1 \label{sec:MT-1} }
The MT-1 is the difficulty of extraction of the decay amplitude
of the two-pion state with non-zero momenta.
Originally Miani and Testa found the problem for the $K\to\pi\pi$ Green's function
in infinite volume Euclidean space-time~\cite{MT-nogo}.
Here we consider a finite-volume Green's function 
given by
\begin{equation}
G_n (t) = \langle 0 | \ (\pi\pi)_n(t) \ {\cal O}(0) \ K(t_K) \ | 0 \rangle
\label{GKPP_1}
\ ,
\end{equation}
where $(\pi\pi)_n(t)$ is the interpolating field
for the $n$-th excited two-pion state
given by $(\pi\pi)_n(t) = \pi(\vec{p}_n,t) \pi(-\vec{p}_n,t)$
and $p_n^2=(2\pi/L)^2\cdot n$, 
${\cal O}(0)$ is the weak operator and $K(t_K)$ is $K$ meson filed.

Inserting the energy eigenstates,
the Green's function (\ref{GKPP_1}) can be rewritten as
\begin{equation}
G_n (t)
= {\sum}_j \ V_{nj} \cdot A_j \cdot \Delta_j(t) \cdot G^{K}(t_K)
\label{GKPP_2}
\ ,
\end{equation}
for the time region $t \gg 0 \gg t_K$,
where $A_j = \langle \overline{(\pi\pi)}_j | {\cal O} | K \rangle$,
$V_{nj} = \langle 0 | (\pi\pi)_n | \overline{(\pi\pi)}_j \rangle$, and
$\Delta_j(t) = \exp ( - \bar{E}_j \cdot t )$.
The state $| \overline{(\pi\pi)}_j \rangle$ is the 
$j$-th two-pion energy eigenstate with the energy $\bar{E}_j$, 
which is different from twice the pion energy $E_j$
due to the two-pion interaction on the finite volume.
The energy $\bar{E}_j$ satisfies the
L\"uscher quantization condition~\cite{Luscher}.
The state $| K \rangle$ is the zero momentum $K$ meson state and
$G^{K}(t) = \langle K | K | 0 \rangle \cdot \exp( - m_K \cdot t_K )$.

The crucial point is 
that $V_{nj} \not\propto \delta_{nj}$ in (\ref{GKPP_2}) generally,
because the interpolating field $(\pi\pi)_n$ has no definite energy
and it can emit the state $| \overline{(\pi\pi)}_j \rangle$ with $j\not=n$.
Thus, $G_{n}(t)$
contains many exponential terms generally.
The extraction of the decay amplitude from such a multi-exponential function
is not trivial.
For $n=2$, for example, 
the dominant contributions of $G_{n}(t)$ for the large time region
come from the states with $j = 0, 1, 2$, 
and the effects from these states disturb extraction of the amplitude $A_2$.
A special case is $n=0$,
for which we can extract the amplitude by a single exponential fitting,
because contaminations from lower states are absent.
%This difficulty of the extraction of the amplitude is
%one of the Maini-Testa problems (the MT-1).
%
%----------------------------------------------------------------------------
%
\section{ Method for avoiding the MT-1 }
There are three methods known for avoiding the MT-1.
These are to calculate the Green's functions for the amplitudes of
(1) $K\to\pi(\vec{0})\pi(\vec{0})$~\cite{BDHS},
(2) $K\to\pi$~\cite{BDSPW}, and 
(3) $K\to\pi(\vec{p})\pi(\vec{0})$ at $p^2=(2\pi/L)^2$ (SPQR kinematics)~\cite{SPQCDR_1}.
As we noted in Sec.~\ref{sec:MT-1},
the MT-1 does not apply to the zero momentum two-pion.
Further this is also true in case (3),
because only one state is dominant for the large time region.
Results of lattice calculations
by the above three methods are reviewed in Sec~\ref{sec:results}.

In the methods (1) and (2),
chiral perturbation theory (CHPT) is used
to reconstruct the physical decay amplitudes from those on the lattice.
Here we consider, as an example, the case
of operators for $(27_L,1_R)$ and $(8_L,1_R)$ representations after subtraction
of the lower dimensional operators.
( See Refs.~\cite{Q78_cCHPT1,Q78_cCHPT2} for $(8_L,8_R)$ operators ).
For simplicity
we set $m_K=m_\pi=M$ and consider the amplitude in the infinite volume.
In this case
the decay amplitude     in the method (1) : $A^{(1)}$
and $K\to\pi$ amplitude in the method (2) : $A^{(2)}$
are given by
\begin{eqnarray}
\lefteqn{
A^{(1)}
= {M^2}/{f^3}\cdot
      \bigl[
           \alpha + \alpha C \cdot \lambda(M)
         + \beta(\Lambda) \cdot M^2 \bigr]
\label{CHPT_1}
,
}
&& \\
\lefteqn{
A^{(2)}
= {M^2}/{f^2}\cdot
      \bigl[
           \alpha + \alpha D \cdot \lambda(M)
         + \gamma(\Lambda) \cdot M^2 \bigr]
\label{CHPT_2}
, 
}
&& 
\end{eqnarray}
in one-loop order of CHPT,
where $f$ is the pion decay constant,
$\lambda(M) = M^2 \cdot \log(M^2/\Lambda^2)$,
and $\Lambda$ is a momentum cut-off for the regularization of CHPT.
The amplitudes for a finite volume
both in quenched and full QCD are presented
in Refs.~\cite{Golterman-Leung_LCHPT,Golterman-Pallante_LCHPT,Golterman-Pallante2_LCHPT}.

The physical decay amplitude is given by
\begin{equation}
A
= ( m_K^2 - m_\pi^2 )/f^3 \cdot
      \bigl[
           \alpha + \alpha E \cdot \lambda(m_K)
         + \delta(\Lambda) \cdot m_K^2 \bigr]
\label{CHPT_3}
,
\end{equation}
where $m_K$ and $m_\pi$ are physical $K$ meson and pion mass.
$m_\pi$ is neglected in the bracket in (\ref{CHPT_3}); 
if this is not made, the structure is very complicated,
which are presented
in Refs.~\cite{Kambor-Missimer-Wyler_cCHPT,Ecker-Kambor-Wyler_cCHPT,Bijnens-Pallante-Prades_cCHPT}.

In (\ref{CHPT_1}--\ref{CHPT_3})
$\alpha$ is a universal constant and independent of kinematics.
The term proportional to $\lambda(M)$ is the ``chiral logarithm term'', and
the associated constants $C$, $D$, and $E$ are determined
by calculation in one-loop order of CHPT.
The third terms are the counter terms for the one-loop order, 
which come from tree order diagram of the next order of the CHPT operators.
The associated constants $\beta$, $\gamma$, and $\delta$
depend on the kinematics and the momentum cut-off $\Lambda$.

In usual lattice calculations,
the universal constant $\alpha$ is obtained by
a chiral extrapolation of the $A^{(1)}$ or $A^{(2)}$
with the fitting function (\ref{CHPT_1}) or (\ref{CHPT_2}),
where $\beta$ and $\gamma$ are unknown constants of the fitting.
Since the precise value of the counter term $\delta$
in the physical amplitude is not known,
it is neglected in usual lattice calculations.
This is the  cause of a large ambiguity 
for the lattice prediction of the amplitude.
Determination of counter terms from lattice calculations
of several amplitudes are discussed 
in Refs.~\cite{Laiho-Soni_CHPT,SPQCDR_LCHPT1,SPQCDR_LCHPT2}.

The chiral logarithm term in the amplitude on the lattice
also depends on the lattice volume and whether calculated 
in quenched or full QCD.
Very recently, Lin {\it et.al.} completed a CHPT calculation
of the $\Delta I=3/2$ decay amplitudes
for $Q_1$, $Q_7$, and $Q_8$ operators
at general kinematics both in quenched and full QCD~\cite{SPQCDR_LCHPT2}.
%
%----------------------------------------------------------------------------
%
\section{ Recent lattice results \label{sec:results} }
%
%In this section recent results obtained by the above three methods
%are reviewed.
%
%-----------------------------
%
\subsection{ Calculation of $K\to\pi(\vec{0})\pi(\vec{0})$ }
JLQCD collaboration calculated ${\rm Re}A_2$
from the $K\to\pi(\vec{0})\pi(\vec{0})$ decay amplitude~\cite{JLQCD_KPP_ReA2}.
They used the plaquette gauge and Wilson fermion action
at $\beta=6.1$ ($1/a=2.67\ {\rm GeV} )$
at $m_K=m_\pi=M$ in the quenched approximation on $24^3$ and $32^3$ lattices.
They reconstructed  the physical decay amplitude $A^{\rm Ph.}$
from those on the lattice $A^{\rm Lat.}$
by one-loop order of CHPT relation
obtained by Golterman and Leung~\cite{Golterman-Leung_LCHPT}.
It is given by
$
A^{\rm Ph.} = A^{\rm Lat.} (m_K,E) \cdot G \cdot [ 1 + M^2 / (16\pi^2f^2) \cdot F(EL/2) ]
$,
where $m_K$ is the $K$ meson mass and $E$ is the two-pion energy
on the lattice,
which equals $m_K=M$ and $E=2M$ in their calculations.
$G$ is the CHPT correction factor in the infinite volume and
the finite volume correction is given by
$F(ML) = 17.827/(ML) + 12\pi^2/(ML)^3$.

Very recently, Lin {\it et.al.} found a misinterpretation
of the $O(1/L^3)$ term in $F(ML)$~\cite{SPQCDR_LCHPT2}.
The amplitude obtained by the lattice calculation
is $A^{\rm L.}(m_K,E+\Delta E)$, and not $A^{\rm L.}(m_K,E)$,
where $\Delta E$ is the energy shift
due to the two-pion interaction on the finite volume which is given by
$\Delta E = 1 / ( 2 f^2 L^3)$ in one-loop order of CHPT.
Thus, in order to use their relation to obtain $A^{\rm Ph.}$,
we should correct the $O(1/L^3)$ term in $F(ML)$ to 
$F(ML) = 17.827/(ML) + 2 \pi^2/(ML)^3$,
which is same as the 
correction factor given by Lellouch and L\"uscher (see Sec.\ref{MT-2.sec}).
JLQCD did not consider this,
but the effect of modification is negligible
in their simulation points.

\begin{figure}[t]
\vspace*{-0.6cm}
\centerline{ \epsfig{ file=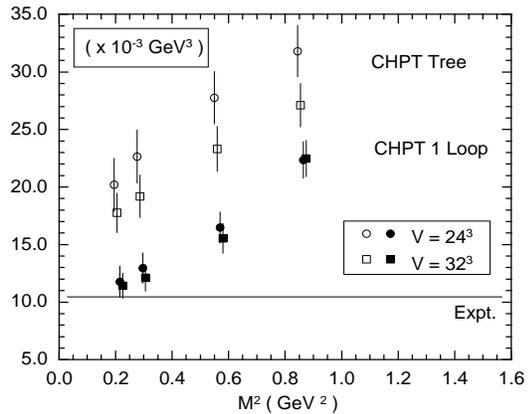, height=5.5cm, width=7.0cm } }
\vspace*{-1.0cm}
\caption{
\label{JLQCD_Kpipi.fig}
JLQCD results of ${\rm Re} A_2$.
}
\vspace*{-0.8cm}
\end{figure}

JLQCD results of ${\rm Re} A_2 \cdot \sqrt{3} / [ G_F V_{us}^* V_{ud} ]$
obtained by tree-level (CHPT Tree) and one-loop order of 
CHPT relation (CHPT 1 Loop )
are plotted in Fig.~\ref{JLQCD_Kpipi.fig}.
Here the counter terms of CHPT for both the lattice and the physical amplitude
are neglected.
%
%----------------------------------------------------
%
%
%----------------------------------------------------
%
We find that
the volume dependence seen with the tree-level analysis
is removed after the finite volume corrections at the one-loop level.
At the same time, the amplitude decreases by $30-40$\%.
Another noteworthy feature
is that a sizable lattice meson mass $M$ dependence still remains
in the amplitude.

Their final result is
${\rm Re} A_2 \cdot \sqrt{3} / [ G_F V_{us}^* V_{ud} ] =
8.9(1.7) \times 10^{-3}\ {\rm GeV}^3 - 11.4(1.5) \times 10^{-3} \ {\rm GeV}^3$,
depending on the choice of the scale for the operator matching
and the CHPT cut-off.
These values are obtained by a chiral extrapolation
of the data in Fig.~\ref{JLQCD_Kpipi.fig}, 
assuming that the remaining mass dependence
comes only from neglecting the counter terms of CHPT on the lattice.
JLQCD results are consistent with 
the experiment $10.4\times 10^{-3}\ {\rm GeV}^3$.
However, a sizable one-loop correction of CHPT
raises the question whether ignoring higher order corrections can be justified.
%
%-------------------------------------------------------------------------------------------------
%
\subsection{ Calculation of $K\to\pi$ }
The $K\to\pi$ amplitudes have been calculated by
Pekurovsky and Kilcup~\cite{PK_KP_Qall}, RBC~\cite{RBC_KP_Qall}, and
CP-PACS collaboration~\cite{CP-PACS_KP_Qall}.
Pekurovsky and Kilcup worked
with the plaquette gauge and the staggered fermions action 
at $\beta=6.0$ ($1/a=2.1\ {\rm GeV}$) and $\beta=6.2$ ($1/a=2.8\ {\rm GeV}$)
in quenched QCD, and $\beta=5.7$ ($1/a=2.0\ {\rm GeV}$) for $N_F=2$ full QCD.
RBC and CP-PACS collaboration employed the 
domain wall fermions in the quenched approximation.
The gauge action is different (RBC : plaquette action, CP-PACS : RG-improved 
action),
but the lattice cut-off is similar ($1/a \sim 2.0\ {\rm GeV}$).
The three groups 
set $m_K=m_\pi=M$ 
for the $K\to\pi$ amplitude.
% for the calculation of the $K\to\pi$ amplitude.
%
%-----------------------------------------------------------------------------------------
%
\subsubsection{ Results for ${\rm Re} A_0$ and ${\rm Re} A_2$ }
\begin{figure}[t]
\vspace*{-0.6cm}
\centerline{ \epsfig{ file=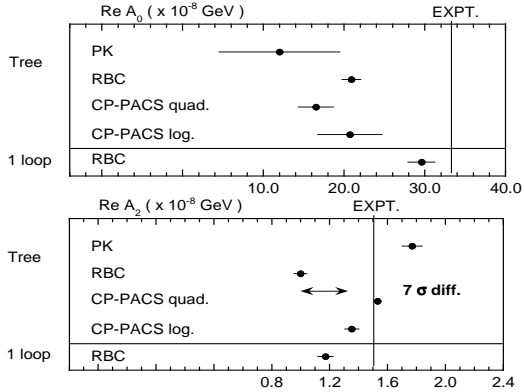, height=5.2cm, width=7.0cm } }
\vspace*{-0.9cm}
\caption{
\label{PK_RBC_CP-PACS.A0A2.fig}
Results for ${\rm Re} A_0$ and ${\rm Re} A_2$
obtained from $K\to\pi$ amplitude.
}
\vspace*{-0.8cm}
\end{figure}

The results for ${\rm Re} A_0$ and ${\rm Re} A_2$
obtained by the three groups are summarized 
in Fig.~\ref{PK_RBC_CP-PACS.A0A2.fig} ; 
PK stands for Pekurovsky and Kilcup, and
``Tree'' and ``1 loop'' refer to the order of CHPT
for the physical decay amplitude.
%
%----------------------------------------------------
%
%
%----------------------------------------------------
%
We should compare the results in the same order of CHPT,
because it is independent of the lattice calculations.
The lattice results for ${\rm Re} A_0$ are almost consistent within 
the three groups, but smaller than the experiment
${\rm Re} A_0 = 33.3\times 10^{-8}\ {\rm GeV}$.
It is also seen that the results for ${\rm Re} A_2$
are inconsistent within the lattice results.
In particular the difference between RBC and CP-PACS calculated
with the same fermion action is about $7\sigma$.

\begin{figure}[t]
\vspace*{-0.6cm}
\centerline{ \epsfig{ file=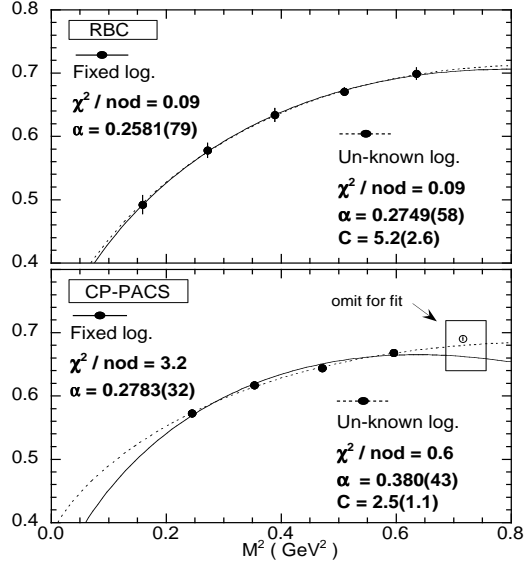, height=7.5cm, width=7.0cm } }
\vspace*{-1.0cm}
\caption{
\label{BK_comp.fig}
Bare $B_K$ obtained from $K\to\pi$ amplitude.
}
\vspace*{-0.8cm}
\end{figure}

The dominant operators for ${\rm Re} A_2$ are the $Q_1$ and $Q_2$.
At $m_K=m_\pi=M$,
the $K\to\pi$ amplitudes for these operators
are related to the value of $B_K$ by
$\langle \pi^+   | Q_1^{(3/2)}   | K^+ \rangle
=\langle \pi^+   | Q_2^{(3/2)}   | K^+ \rangle
=B_K \cdot 9/4 \cdot (M f_K)^2$,
where $Q_j^{(3/2)}$ is the $\Delta I=3/2$ part of the operator $Q_j$.
Thus, ${\rm Re}A_2$ is proportional
to $B_K$ in the chiral limit approximately.
In Fig.~\ref{BK_comp.fig}
we show the $M$ dependence for the bare $B_K$
obtained from $K\to\pi$ amplitude by RBC and CP-PACS.
Here RBC results are those obtained by the wall-wall normalization
which is used for the estimation of ${\rm Re}A_2$.
%
%----------------------------------------------------
%
%
%----------------------------------------------------
%

We attempt to fit the  $B_K$ data
with the function predicted by CHPT :
$B_K = \alpha \cdot ( 1 - C\cdot L(M) ) + \beta\cdot M^2$,
where $L(M)=1/(16\pi^2 f^2) \cdot M^2 \cdot \log(M^2)$.
The coefficient of the chiral logarithm term $C$ is dealt with in two ways :
fixed constant at the CHPT prediction $C=6$ (Fixed logarithm) or 
an unknown constant (Unknown logarithm).
The CP-PACS data at the largest $M$,
enclosed by box in the figure, is omitted in the fitting.
In Fig.~\ref{BK_comp.fig}
the results of the fitting are tabulated.
The fitting curves are also plotted by solid lines 
for the ``Fixed logarithm'' fitting
and broken lines for the ``Unknown logarithm'' fitting.
The $M$ dependence of the RBC data are consistent with the CHPT prediction,
but those of the CP-PACS are not.
The value of the chiral logarithm given by ``Unknown logarithm'' fitting
is inconsistent with the CHPT prediction $C=6$.
This inconsistency with CHPT
is also found in previous CP-PACS work of $B_K$~\cite{CP-PACS.BK}.
In their calculations of the $K\to\pi\pi$ decay amplitudes,
the final results are evaluated
by the chiral extrapolation
with the quadratic polynomial function and the ``Unknown logarithm'',
which are refereed to as 
"CP-PACS quad." and "CP-PACS log."
in Fig.~\ref{PK_RBC_CP-PACS.A0A2.fig}, respectively.
The final results of RBC are given by the ``Fixed logarithm'' fitting.

This inconsistency of the $M$ dependence between the two groups
is reflected in the large discrepancy of the final results of ${\rm Re}A_2$.
It should be noted that the renormalization factor
can not change the $M$ dependence.
A possible reason for of this is the lattice cut-off error
because the gauge actions adopted by the two groups are different.
Numerical investigations on the mass dependence of $B_K$
closer to the continuum limit are necessary
to consolidate calculations of the $K\to\pi\pi$ decay amplitude.
%
%-----------------------------------------------------------------------------------------
%
\subsubsection{ Results for $\epsilon'/\epsilon$ }
We found an inconsistency within the lattice results for ${\rm Re} A_2$. 
The lattice results are also troublesome for $\epsilon'/\epsilon$.  
As shown in Table.~1 they are far from the experimental values
obtained by KTeV~\cite{KTeV_EPE} and NA48 groups~\cite{NA48_EPE}.
%
%----------------------------------------------------
%
\begin{table}[t]
\vspace*{-0.6cm}
\begin{center}
\begin{tabular}{lll}
\hline
\hline
PK       &  $ - 38.6 \pm 2.1\pm 9.1$ & Tree       \\
RBC      &  $ - 3.2  \pm 2.2$        & Tree       \\
         &  $ - 4.0  \pm 2.3$        & One-loop   \\
CP-PACS  &  $ - 7.7  \pm 2.0$        & Tree       \\
\hline
EXPT     &  $ + 20.7 \pm 2.8$        & KTeV       \\
         &  $ + 15.3 \pm 2.6$        & NA48       \\
\hline
\hline
\end{tabular}
\end{center}
\vspace*{-0.1cm}
\label{EpsilonP_Epsilon.table}
Table 1. Results of $\epsilon'/\epsilon$ ( $\times 10^{-4}$ ).
\vspace*{-0.9cm}
\end{table}
%
%----------------------------------------------------
%

One of possible reasons for the discrepancy is the quenched approximation.
The dominant operators for $\epsilon'/\epsilon$ are $Q_6$ and $Q_8$.
Golterman and Pallante pointed out that
the naive relation between the $K\to\pi$
and the $K\to\pi\pi$ decay amplitude for $Q_6$ operator
shown in (\ref{CHPT_2}) and (\ref{CHPT_3}) are broken in the quenched QCD
at $O(M^2)$~\cite{Golterman-Pallante_Q6}.
As they discussed, 
this problem can be avoided by removing the contraction
between $q$ and $\bar{q}$
in $Q_6 = \sum_{q=u,d,s} (\bar{s}_a d_b )_{L_\mu} (\bar{q}_b q_a )_{R_\mu}$.
(An alternative method is discussed in Ref.~\cite{Golterman-Pallante2_Q6}).
Bhattacharya {\it et.al.} investigated the effect of removing the contractions
and found the effect to be very large for the $Q_6$ operator
(about $20 - 100\%$)~\cite{Batacharya-ETAL_Q6}.
CP-PACS collaboration also examined the effect for $\epsilon'/\epsilon$.
They find that while the effect is very large, 
the value of $\epsilon'/\epsilon$ is still negative
$\epsilon'/\epsilon = -1.70(53)\times 10^{-4}$~\cite{CP-PACS_Q6}.

Another possible reason is the lack of final state interactions
in the $K\to\pi$ amplitude, which are 
considered to be important for the $\Delta I=1/2$ process.
Exploring methods for a direct calculation of the decay amplitude
from the $K\to\pi\pi$ Green's function is strongly desirable.
%
%-------------------------------------------------------------------------------------------------
%
\subsection{ Calculation at SPQR kinematics }
SPQCDR collaboration calculated the $K\to\pi\pi$ decay amplitude
from the $K\to\pi\pi$ Green's function
at SPQR kinematics directly~\cite{SPQCDR_1,SPQCDR2_calc}.
In this section we show their preliminary results~\cite{SPQCDR2_calc}.

They calculate the $\Delta I=3/2$ amplitude
for $Q_1$, $Q_7$, and $Q_8$
which are the dominant operators for ${\rm Re A}_2$
and the $\Delta I=3/2$ part of $\epsilon'/\epsilon$.
The calculations are carried out with the plaquette gauge
and the Clover-Wilson fermion action
with non-perturbatively determined $C_{SW}$
at $\beta=6.0$ on a $24^3\times 64$ lattice.
The amplitudes for 
$\pi(\vec{p})\pi(\vec{0})$ at $p^2=0$ and $(2\pi/L)^2$
are converted to the physical one
by the one-loop order of CHPT relations obtained in Ref.~\cite{SPQCDR_LCHPT2}.
Finite volume effects have not been included in the preliminary results.
%
%----------------------------------------------------
%
\begin{figure}[t]
\vspace*{-0.6cm}
\centerline{ \epsfig{ file=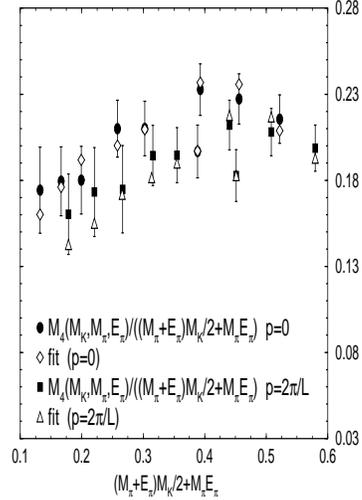, height=7.0cm, width=4.7cm, angle=-90 } }
\vspace*{-0.9cm}
\caption{
\label{Papinutto.O4.fig}
SPQCDR results of $M_4$.
}
\vspace*{-0.8cm}
\end{figure}
%
%----------------------------------------------------
%

In Fig.~\ref{Papinutto.O4.fig}
the ratio between the amplitude $M_4 \equiv \langle \pi^+\pi^0 | Q_1 | K^+ \rangle$
and the kinematic constant of the lowest order of CHPT
( $( M_\pi + E_\pi )\cdot M_K/2 + M_\pi\cdot E_\pi$ )
is plotted.
Here $M_\pi$ and $M_K$ are pion and $K$ meson masses,
and $E_\pi=\sqrt{M_\pi^2+p^2}$.
The figure shows that the ratio depends on the kinematics.
This means that higher order effects of CHPT are significant.
Their fit results with the one-loop order of CHPT relation are also plotted
by open symbols.

The preliminary results for the physical amplitudes are
$M_4 = 0.0135(80)\ {\rm GeV}^3$ (the experiment $=0.0104\ {\rm GeV}^3$),
$\langle Q_7 ( 2\ {\rm GeV}) \rangle = 0.14(2)\ {\rm GeV}^3$, and
$\langle Q_7 ( 2\ {\rm GeV}) \rangle = 0.71(6)\ {\rm GeV}^3$.
They also found that the effects of the one-loop order of CHPT
are large, $65\%$ for $Q_1$, $-34\%$ for $Q_7$, and $-24\%$ for $Q_8$
in their kinematic range.
A sizable one-loop correction
raises the question whether ignoring higher order corrections can be justified.
%
%----------------------------------------------------------------------------
%
\section{ New ideas to solve MT-1 }
\subsection{ Diagonalization method }
The problem of MT-1 also appears in the pion 4-point function given by
$g_{nm}(t) = \langle 0 | (\pi\pi)_n (t)\ (\pi\pi)_m (0) | 0 \rangle$,
where the same definition as in (\ref{GKPP_1}) is used.
For large time regions $t \gg 0$ the 4-point function behaves as
$g_{nm}(t) = {\sum}_j V_{nj} \cdot \Delta_j (t) \cdot V^{T}_{jm}$,
where
$V_{nj}=\langle 0 | (\pi\pi)_n | \overline{(\pi\pi)}_j \rangle$,
$\Delta_j(t) = \exp ( - \bar{E}_j \cdot t )$, and
$| \overline{(\pi\pi)}_j \rangle$ is the energy eigenstate
with energy $\bar{E}_j$.
We realize that
the pion 4-point function contains many exponential terms
similar to the $K\to\pi\pi$ Green's function.

This problem can be solved by a diagonalization of the matrix
$M(t,t_0) = g^{-1/2}(t_0)$ $g(t)$ $g^{-1/2}(t_0)$
at each $t$,
where $t_0$ is some reference time
and the momenta $n$ and $m$ are regarded as matrix indices.
The eigenvalues of $M(t,t_0)$ take the form 
$\lambda_j(t,t_0) = \exp ( - \bar{E}_j \cdot [ t - t_0 ] )$,
and
$
V_{nj}=\langle 0 | (\pi\pi)_n | \overline{(\pi\pi)}_j \rangle
= [ g^{-1/2}(t_0)\ U(t_0)\ \Delta^{-1/2}(t_0)]_{nj}
$,
where $U(t_0)$ is an orthogonal transformation matrix
for the diagonalization of $M(t,t_0)$.

If the matrix $V$ can be determined precisely from the study
of the two-pion system,
we can solve the MT-1 for the $K\to\pi\pi$ Green's function by 
defining a new function 
$
\overline{G}_n (t) \equiv {\sum}_j V^{-1}_{nj} \cdot G_j (t)
= A_n \cdot \Delta_n(t) \cdot G^K(t_K)
$,
where $G_j(t)$ is the $K\to\pi\pi$ Green's function defined by (\ref{GKPP_1}), and
$A_n = \langle \overline{(\pi\pi)}_n | {\cal O} | K \rangle$.
We expect that
the new Green's function $\overline{G}_n (t)$
behaves as a single exponential function for large times
and the decay amplitude $A_n$ can be extracted from it easily.

This diagonalization method was proposed by 
L\"uscher and Wolf~\cite{Luscher-Wolff}. 
It has been applied to 
many statistical systems~\cite{Luscher-Wolff,Luscher-Wolf_stat,Rummukainen-Gottlib}
and also to the $I=2$ two-pion system in QCD
by Fiebig {\it et.al.}~\cite{Fiebig}
and by CP-PACS collaboration~\cite{CP-PACS.PHSH}.
In particular CP-PACS evaluated the $I=2$ pion phase shift
with small statistical errors, which is extracted 
from the energy $\bar{E}_j$ obtained by the lattice simulation
using L\"uscher's quantization condition. 

The CP-PACS calculations are carried out
with the plaquette gauge and the Wilson fermion action 
in the quenched approximation at $\beta =5.9$ ($1/a=1.9\ {\rm GeV}$)
on three lattice volumes ($24^3$, $32^3$, and $48^3$).
Since the entire matrix of the $4$-point function $g_{nm}(t)$ cannot be obtained
in actual lattice calculations, they set a momentum cut-off 
$p^2_{\rm cut} = (2\pi/L)^2\cdot N$
and obtain the eigenvalues $\lambda_n(t)$ for $n\leq N$.
The cut-off dependence is also investigated.

In order to examine the effects of diagonalization,
it is convenient to consider two ratios,
$R_n(t) = g_{nn}(t)         / [g_n^\pi(t)]^2$ and
$D_n(t) = \lambda_n (t,t_0) / [g_n^\pi(t)]^2$,
where $g_n^\pi(t)$ is the pion $2$-point function
with momentum $p_n^2=(2\pi/L)^2\cdot n$.
If the $4$-point function contains only a single exponential term, then
$R_{n}(t) = \exp [ - \Delta E_n \cdot t ]$,
where $\Delta E_n \equiv \bar{E}_n - E_n$ and $E_n$
is twice the $n$-th pion energy.
If the momentum cut-off is large enough, then
$D_n(t) = \exp [  - \Delta E_n \cdot ( t - t_0 ) ]$.

\begin{figure}[t]
\vspace*{-0.6cm}
\centerline{ \epsfig{ file=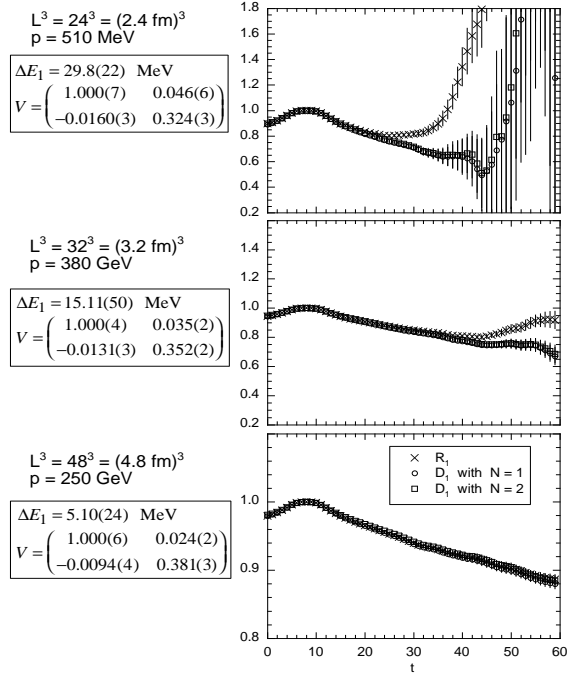, height=9.0cm, width=7.5cm } }
\vspace*{-1.0cm}
\caption{
\label{CP-PACS.PHSH.fig}
CP-PACS results of two ratio $R_n(t)$ and $D_n(t)$ for $n=1$.
}
\vspace*{-0.9cm}
\end{figure}

In Fig.~\ref{CP-PACS.PHSH.fig}
the two ratios $R_n(t)$ and $D_n(t)$
for $n=1$ at $m_\pi/m_\rho=0.692$ are plotted.
%
%----------------------------------------------------
%
%
%----------------------------------------------------
%
The pion source is located at $t=8$.
The momentum cut-off is set at $N=1$ and $N=2$.
It is found that the diagonalization is effective for the small lattice volume
while it is not for the large volumes in the figure.
The momentum cut-off dependence is negligible, however.
A single exponential behavior can be seen for the ratio after 
diagonalization $D_n(t)$ for all volumes.
The values of the energy shift $\Delta E_n$ and
$V_{nj}=\langle 0 | (\pi\pi)_n | \overline{(\pi\pi)}_j \rangle$
for the cut-off $N=1$
are also tabulated in Fig.~\ref{CP-PACS.PHSH.fig}.
That these are obtained with small statistical errors 
demonstrate that the MT-1 can be solved by the diagonalization method.
Furthermore their results for the $48^3$ lattice 
show that the MT-1 is not serious  for $n=1$ on this lattice.
Of course the CP-PACS investigation is limited to the $I=2$ two-pion system; 
the study of the $I=0$ system is necessary for the $\Delta I=1/2$ decay process.

Finally we comment on the quenched approximation.
It is known from quenched CHPT
that serious problems appear in the amplitude for 
the two-pion system in quenched QCD
due to lack of unitarity~\cite{Bernard-Golterman,Colangelo-Pallante} :
Chiral and un-physical threshold divergences appear 
for both $I=0$ and $I=2$ two-pion systems,
and enhanced finite volume effects for $I=0$ are present.
These problems are also predicted to appear
in the $K\to\pi\pi$ decay amplitudes~\cite{Golterman-Pallante2_LCHPT,SPQCDR_LCHPT2,SPQCDR_LCHPT4,Pallante_LCHPT}.
While the presence such pathologies has not been numerically confirmed 
in actual lattice simulations, 
we should make our study in full QCD in order to avoid the uncontrollable 
quenching problems. 
%
%----------------------------------------------------------------------------
%
\subsection{ $G$-periodic boundary condition }
At this conference Kim~\cite{Kim} proposed a new idea of 
imposing the $G$-periodic boundary condition for $u$ and $d$ quark
and $C$-periodic boundary condition for $s$ quark in the $z$-direction.
The explicit forms are given by
$u(x+\hat{z}\cdot L)=-i \cdot C \bar{d}^T(x)$,
$d(x+\hat{z}\cdot L)=+i \cdot C \bar{u}^T(x)$, and
$s(x+\hat{z}\cdot L)=         C \bar{s}^T(x)$, where
$C$ is the charge conjugation matrix ($C^{-1}\gamma_\mu C = \gamma_\mu^T$).
The periodic boundary condition is imposed in other directions for all quarks.
Since pion is an odd eigenstate of $G$-parity, 
the boundary condition for pion in the $z$-direction is anti-periodic
and allowed $z$-components of momentum are restricted to 
$p_z=\pi/L \cdot ( 2j + 1 )$ ($j=0,1,\cdots$).

Kim proposed to extract the decay amplitude from
$G(t)=\langle 0 | (\pi\pi)(t) {\cal O}(0) K(t_K) | 0 \rangle$,
where $(\pi\pi)(t) = \pi(\vec{p},t) \pi(-\vec{p},t)$ and $\vec{p}=\hat{z}\pi/L$.
Since this operator is the interpolating operator
for the lowest two-pion state under the $G$-periodic boundary condition,
$G(t)$ behave as a single exponential function for large $t$, 
and the decay amplitude should be extracted easily.
Numerical implementation of this method is future work.
%
%----------------------------------------------------------------------------
%
\section{ The MT-2 \label{MT-2.sec} }
We finally consider the second part of the Maiani-Test problem (the MT-2).
As we explained above, 
the MT-1 can be avoided or solved by appropriate methods.
However, even if we succeed in the extraction of the $K\to\pi\pi$ decay amplitude
$A^{\rm Lat.} = \langle \overline{(\pi\pi)} | {\cal O} | K \rangle$
from the $K\to\pi\pi$ Green's function
by lattice simulations, another problem remains.
The lattice amplitude $A^{\rm Lat.}$ is that
for the two-pion energy eigenstate on a finite Euclidean space-time, 
and not in the infinite volume Minkowski space-time.
We can relate $A^{\rm Lat.}$ to the physical amplitude $A^{\rm Ph.}$
by using some effective theory, 
but using such theories is the cause of large uncertainties 
in the lattice prediction of the decay amplitude
as we discussed in the previous sections.
Recently Lellouch and L\"uscher
obtained a relation between $A^{\rm Lat.}$ and $A^{\rm Ph.}$
at two-pion energy $\bar{E} = m_K$~\cite{Lellouch-Luscher}.
Here we show a brief derivation.

The two-pion energy $\bar{E}$ on the lattice
satisfies the L\"uscher quantization condition~\cite{Luscher}.
Introducing a ``weak'' interaction $H^W=\pi\pi K \cdot A^{\rm Lat.}$, 
the energy is further shifted from $\bar{E}$ to $\bar{E}'$
on finite volumes,
which also satisfies the L\"uscher quantization condition,
{\it i.e.}
\begin{equation}
\delta  (\bar{E} ) =  - \phi(\bar{E} ) \ , \
\delta' (\bar{E}') =  - \phi(\bar{E}') \ \mbox{ $mod$ $\pi$}
\label{LL_1}
\ ,
\end{equation}
where
$\delta (\bar{E}) $ is the pion phase shift for the strong interaction, and
$\delta'(\bar{E}')$ is that for the strong and the weak interactions.
The spherical zeta function $\phi$ is common for both cases.

The energy shift $\bar{E}' - \bar{E}$ can be estimated by perturbation theory
for the weak interaction on a finite volume.
In case of $\bar{E}=m_K$, one finds 
\begin{equation}
\bar{E}' - \bar{E} = \pm | A^{\rm Lat.}(\bar{E})|
\label{LL_2}
\ ,
\end{equation}
where it should be noted that the energy shift is first order in 
the weak interaction. This is because the energy is degenerate at $H^W=0$.
We can also estimate the difference of the phase shifts at $\bar{E}'$
by perturbation theory in the infinite volume, which yields
\begin{equation}
\delta'(\bar{E'}) - \delta(\bar{E}')
= \frac{1}{8\pi} \cdot \frac{ \bar{p}' }{ \bar{E}' } \cdot
\frac{ |A^{\rm Ph.}(\bar{E'})|^2 }{ P_K^2 + m_K^2 }
\label{LL_3}
\ ,
\end{equation}
where $\bar{p}' = \sqrt{ \bar{E}'^2 / 4 - m_\pi^2}$,
and the denominator is the $S$-channel $K$ meson propagator
with 4-momentum $P_K=(i\bar{E'},\vec{0})$.
Substituting (\ref{LL_3}) into the L\"uscher quantization condition (\ref{LL_1}),
and expanding for $\bar{E}' - \bar{E}$ by using (\ref{LL_2}),
we obtain the following formula at $\bar{E}=m_K$
(LL-formula).
\begin{equation}
| A^{\rm Ph.}(\bar{E})|^2
= C(\bar{E}) \cdot \rho(\bar{E}) \cdot | A^{\rm Lat.}(\bar{E})|^2
\label{LL_F}
\ ,
\end{equation}
where $C(m_K) = 32\pi^2 \cdot m_K^2 / \bar{p}$,
$\bar{p} = \sqrt{ \bar{E}^2 / 4 - m_\pi^2}$, and
$\rho(\bar{E}) =
\bar{E} / (4\pi \bar{p} ) \cdot
 {\rm d}/ ( {\rm d} \bar{p} ) [ \delta(\bar{E}) + \phi(\bar{E} ) ]
$.

The validity of the LL-formula for the $\Delta I=3/2$ amplitude
is confirmed in one-loop order of CHPT
even in the quenched theory~\cite{SPQCDR_LCHPT2}.
This is not the case for the $\Delta I=1/2$ amplitude.
Since the correction factor depends on the weak operators
due to the lack of unitarity in quenched QCD,
the simple relation from $A^{\rm Lat.}$ to $A^{\rm Ph.}$
can not be obtained~\cite{SPQCDR_LCHPT2,SPQCDR_LCHPT4,Pallante_LCHPT}.

It should be noted that the LL-formula given by (\ref{LL_F})
is relied on the L\"uscher quantization condition
which holds for the energy on a finite periodic box in the two-pion center 
of mass frame.
The formula is not applicable to the SPQR kinematics which is 
in the laboratory frame.
An extension of the LL-formula to the general frame can be obtained easily
using the Rummukaine-Gottlieb quantization condition
for the general frame~\cite{Rummukainen-Gottlib}.

We note that the formula can not be applied for the $G$-periodic boundary condition, 
because the anti-periodic boundary is imposed in the $z$-direction.
Extending the formula to this case is not trivial,
since the condition on the energy in such a boundary condition is not known. 

Lin {\it et al.} derived the LL-formula from a different approach~\cite{LMST}.
They found the following simple relation between 
the two-pion state $| \pi\pi \rangle$ in the physical decay amplitude
and that on the lattice $| \overline{(\pi\pi)} \rangle$ 
at general energy $\bar{E}$ :
$
| (\pi\pi) \rangle  \Leftrightarrow
(4\pi) \cdot \sqrt{ \rho(\bar{E}) \cdot \bar{E} / \bar{p} } \cdot
| \overline{(\pi\pi)} \rangle
$.
This gives
the extension of the LL-formula
to the case of $\bar{E}\not= m_K$.

They also investigated the validity of the LL-formula, and 
found that the volume has to be sufficiently large or the two-pion interaction 
has to be weak enough,
so that the boundary condition does not distort the two-pion wave function.
This is also required for the L\"uscher quantization condition.

At present there are few studies of the two-pion system, and
informations are scarce of the two-pion wave functions.
It is extremely important to investigate the two-pion system,
before embarking on a numerical application of the LL-formula to 
the $K\to\pi\pi$ decay amplitude.
%
%-----------------------------------------------------------------------------------
%
\section{ Summary }
In this article we discussed recent theoretical and numerical progresses
in the calculation of the $K\to\pi\pi$ decay amplitude.
While this calculation has been plagued with a number of difficulties, 
we discussed that most of them had been shown to be theoretically solvable. 
In particular, the MT-1 can be avoided by a judicious choice of kinematics and boundary conditions 
or can be solved by diagonalizations, in principle.

In our opinion, the most desirable way of calculating the decay amplitude is 
the following.
We extract the decay amplitude from the $K\to\pi\pi$ Green's function directly
using the SPQCDR method or one of new ideas described in previous sections.
We then reconstruct the physical decay amplitude from that on the lattice
by the LL-formula or its extensions without using effective theories of QCD.

In this procedure,
a quantitative study of the two-pion system is necessary
to examine the validity of the LL-formula.
Also all calculations should be made in full QCD
to avoid uncontrollable quenching problems
due to the lack of unitarity in the quenched QCD.

The final obstacle to avoid using effective theories 
in the actual lattice calculations is chiral extrapolation. 
Practical lattice simulations
are carried at $m_\pi \sim ( 400 - 800 )\ {\rm MeV}$.
We have to extrapolate our decay amplitude obtained by the LL-formula
to that at the physical mass with some assumption for the mass dependence.
Here effective theories such as CHPT is used.
This causes large uncertainties in the lattice prediction of the decay amplitude,
because there is no reliable effective theory for such a long extrapolation.
A unique solution to this problem is a simulation
at or near the physical point.
%
%----------------------------------------------------------------------------
%
%\vspace{-0.4cm}
\section*{ Acknowledgments }
%\vspace{-0.2cm}
%
I thank particularly
A. Soni, C.T. Sachrajda, A. Vladikas, C. Kim, N. Chirist, M. Golterman,
E. Pllante, and M. Papinutto
for sharing their results with me.
I am also indebted to all members of the CP-PACS and JLQCD collaboration.
%
%============================================================================
%

%
%=============================================================================
%
\end{document}